\documentclass[12pt]{article}
\usepackage{amsfonts}
\usepackage{amssymb,amsmath,amsthm,latexsym}
\usepackage[numbers,compress]{natbib}
\usepackage{amsmath}
\usepackage{cases}
\usepackage{indentfirst}
\usepackage{mathrsfs}
\usepackage{amsfonts}
\usepackage{amsmath}
\usepackage{amsthm}
\usepackage[usenames]{color}
\usepackage{footnote}
\usepackage{longtable}
\usepackage{hyperref}

\allowdisplaybreaks[4]
\newtheorem{theorem}{Theorem}[section]

\newtheorem{lemma}[theorem]{Lemma}
\newtheorem{definition}[theorem]{Definition}

\newcommand{\proved}{{\hfill \bf \fbox{$ $}}}
\newcommand{\F}{\mathbb{F}}

\begin{document}
\title
{\bf Some classes of power functions with low $c$-differential uniformity over finite fields}
\author{Zhengbang Zha$^{1}$, Lei Hu$^{2,3}$\\
\vspace*{0.0cm}\\
{\small $^1$School of Mathematical Sciences, Luoyang Normal University, Luoyang 471934, China}\\
{\small $^2$State Key Laboratory of Information Security, Institute
of Information Engineering,}\\ {\small  Chinese Academy of
Sciences, Beijing 100093, China}\\
{\small $^3$School of Cyber Security, University of Chinese Academy of Sciences, Beijing 100049, China}\\
{\small E-mail: zhazhengbang@163.com; hulei@iie.ac.cn}}
\date{}
\maketitle{}\baselineskip=20pt

\begin{abstract}
Functions with low $c$-differential uniformity have optimal resistance to some types of differential cryptanalysis.
In this paper, we investigate the $c$-differential uniformity of power functions over finite fields. Based on some
known almost perfect nonlinear functions, we present several classes of power functions $f(x)=x^d$ with $_{c}\Delta_f\leq3$.
Especially, two new classes of perfect $c$-nonlinear power functions are proposed.

\vskip 2mm



\end{abstract}

\section{Introduction}

Let $p$ be a prime and $n$ a positive integer. Define $\F_{p^n}$ be a finite field with $p^n$ elements.
Denote $\F_{p^n}\setminus\{0\}$ by $\F_{p^n}^*$. Many ciphers are designed by the functions defined on $\F_{p^n}$.
Differential cryptanalysis is an effective cryptanalytic tool of symmetrical cipher systems \cite{BS}.
For a vectorial Boolean function $f$, Borisov et al \cite{BCJW} proposed a new type of differential $(f(cx),f(x))$ by using modular multiplication.
It was used to cryptanalyze some known ciphers such as a variant of the IDEA cipher. Based on the mentioned work, Ellingsen et al \cite{EFRST} defined
a new multiplicative differential, and presented the relative concept of $c$-differential uniformity as follows.

\begin{definition}[\cite{EFRST}]\label{def1} Let $a,c\in\F_{p^n}$. For a function $f: \F_{p^n}\longrightarrow \F_{p^n}$, the (multiplicative)
$c$ derivative of $f$ with respect to $a$ is defined as
$$_{c}D_{a} f(x)=f(x+a)-cf(x), {\rm for\; all\;} x.$$
Let $_{c}\Delta_{f}(a,b)$ denote the number of solutions $x\in \F_{p^n}$ of $f(x+a)-f(x)=b$, where
$a,b\in \F_{p^n}$. We call $_{c}\Delta_f=\max\{_{c}\Delta_{f}(a,b)|a,b\in
\F_{p^n}, {\rm and}\; a\neq 0\; {\rm if}\; c=1\}$ the $c$-differential uniformity of $f$. If $_{c}\Delta_f=\delta$, then we say that $f$
is differentially $(c,\delta)$-uniform.
\end{definition}

The function $f$ is called perfect $c$-nonlinear (PcN) if $_{c}\Delta_f=1$, and is called almost perfect $c$-nonlinear (APcN)
if $_{c}\Delta_f=2$. Note that the $c$-differential uniformity is equivalent to the usual differential uniformity when $c=1$ and $a\neq0$.
In this case, the PcN and APcN functions are called perfect nonlinear (PN) and almost perfect nonlinear (APN) functions respectively.
When $c=0$ or $c\neq1$ and $a=0$, we have $_{c}D_{a} f(x)=f(x+a)$ or $(1-c)f(x)$ correspondingly, which is linear equivalent to $f(x)$.

In \cite{EFRST}, the authors investigated the $c$-differential uniformity of the $p$-ary Gold PN function and the inverse function. Riera and
St\u{a}nic\u{a} \cite{RS} continued the work, and characterized the $c$-differential uniformity of the Gold function over $\F_{2^n}$. Moreover, they exhibited
the $c$-differential uniformity of some known APN power functions in odd characteristic finite fields. Yan, Mesnager and Zhou \cite{YMZ} completely determined
the $c$-differential uniformity of power functions with Gold exponents, which confirms a conjecture proposed by Bartoli and Timpanella \cite{BT}.
Furthermore, they presented some new classes of power functions with low $c$-differential uniformity.
By using the first kind of Dickson polynomials \cite{HMSY}, Hasan et al \cite{HPRS} introduced some classes of power maps with low $c$-differential uniformity for $c=-1$.
Some PcN power maps over finite fields of odd characteristic were exhibited. Based on the $c$-differential notion, St\u{a}nic\u{a} et al \cite{SGGRT} proposed
the concept of $c$-differential bent functions, and revealed the link between $c$-differential bent and perfect $c$-nonlinear. They presented some constructions
of $c$-differential bent functions including the PcN functions. Recently, Bartoli and Calderini \cite{BC} showed that a quadratic polynomial $f$ is PcN or APcN
if and only if $f$ is a permutation polynomial or 2-to-1 mapping on $\F_{p^n}$ for $c\in\F_p\setminus\{1\}$ respectively. By applying the Akbary-Ghioca-Wang criterion \cite{AGW}
and its generalization \cite{MQ}, they offered several constructions of PcN and APcN polynomial functions.

In this paper, we explore the $c$-differential uniformity of some known APN functions over finite fields.
By calculating the solutions of certain equations over finite fields, we present several classes of
functions with low $c$-differential uniformity.

\section{Preliminaries}

In what follows, we always let $p$ be an odd prime, $n,m,k,d$ be positive integers. Let $\chi$ be the quadratic character of $\mathbb{F}_{p^n}^*$, i.e., $\chi(x)=x^{\frac{p^n-1}{2}}$ for $x\in\mathbb{F}_{p^n}^*$.
If $x$ is a square in $\mathbb{F}_{p^n}^*$, $\chi(x)=1$. Otherwise, $\chi(x)=-1$.

For a $x\in\F_{p^n}$, it can be denoted by $\eta+\eta^{-1}$, where $\eta\in\F_{p^{2n}}^*$.
As $x=\eta+\eta^{-1}=(\eta+\eta^{-1})^{p^n}$, one can get
$$(\eta^{p^n+1}-1)(\eta^{p^n-1}-1)=0,$$
which implies that $\eta\in\mathbb{F}_{p^n}^*$ or $\eta^{p^n+1}=1$.

Below we list some of the known APN power functions \cite{DMMP,HRS,Led,ZW,ZW1} needed later.

\begin{theorem} \label{thm1} Let $f(x)=x^d$ be a function over $\F_{p^n}$. Then $f$ is an APN function if:

(1) $d=3$ and $p>3$;

(2) $d=\frac{2p^n-1}{3}$ and $p^n\equiv2 \pmod{3}$;

(3) $d=\frac{p^n+1}{4}$ and $p^n\equiv7 \pmod{8}$;

(4) $d=\frac{p^n+1}{4}+\frac{p^n-1}{2}$ and $p^n\equiv3 \pmod{8}$;

(5) $p=3$, $d$ is even with $(3^m+1)d-2=k(3^n-1)$, where $k$ is odd, $m$ is a positive integer satisfying that $\gcd(m,n)=1$ and $2m<n$;

(6) $p=5$, $d$ is odd with $(5^m+1)d-2=k(5^n-1)$, where $k$ is odd, $m$ is even such that $\gcd(m,n)=1$.
\end{theorem}

According to the definition of the $c$-differential uniformity, it was shown in \cite{HPRS} that the power functions $x^d$ and $x^{dp^j}$ ($j\in\{0,1,\cdots n-1\}$)
have the same $c$-differential uniformity over $\F_{p^n}$. If $c=\pm1$ and $\gcd(d,p^n-1)$, then the $c$-differential uniformity of $x^d$ and $x^{d^{-1}}$ are the same
over $\F_{p^n}$, where $d^{-1}$ is the inverse of $d$ modulo $p^n-1$. Therefore, if two power functions are Carlet-Charpin-Zinoviev equivalent \cite{CCZ,Dem}, then they
have the same $c$-differential uniformity for $c=\pm1$.

The following lemmas will be used in the sequel.

\begin{lemma} \label{lema} \cite{YMZ} Let $f(x)=x^d$ be a power function over $\F_{p^n}$. If $c=1$, then $_{c}\Delta_f=\max\{_{c}\Delta_{f}(1,b)|b\in
\F_{p^n}\}$. If $c=0$, then $_{c}\Delta_f=\gcd(d,p^n-1)$. Otherwise,
$$_{c}\Delta_f=\max\big\{ \{_{c}\Delta_{f}(1,b)|b\in \F_{p^n}\} \cup \{\gcd(d,p^n-1)\} \big\}.$$
\end{lemma}

\begin{lemma} \label{lemb} \cite{EFRST} Let $k,m,n$ be positive integers. Then
$$
  \gcd(m^k+1,m^n-1)=\left\{
  \begin{array}{cl}
    \frac{2^{\gcd(2k,n)}-1}{2^{\gcd(k,n)}-1}, &{\rm if} \ m=2,\\
    2, &{\rm if} \ m>2, \frac{n}{\gcd(n,k)} \ {\rm is \ odd},\\
    m^{\gcd(k,n)}+1, &{\rm if} \ m>2, \frac{n}{\gcd(n,k)} \ {\rm is \ even}.
  \end{array}
\right.
$$
\end{lemma}

\begin{lemma} \label{lemc} \cite{DMMP} The polynomial $x^{p^k}+bx\in\F_{p^n}[x]$ describes a bijective linear mapping if and only if
$-b$ is not a $(p^k-1)$th power in $\F_{p^n}$.
\end{lemma}

\section{New results on $c$-differential uniformity}

In this section, by utilizing the quadratic character of $\mathbb{F}_{p^n}^*$, we study the $c$-differential uniformity of the APN power functions in Theorem \ref{thm1}.
Six classes of power functions with low $c$-differential uniformity are presented. Two of them are proved to be PcN.

\begin{theorem} \label{thmf} Let $n, k, d$ be odd such that $\gcd(n,k)=1$ and $\frac{3^k+1}{2}\cdot d\equiv\frac{3^n+1}{2}\pmod{3^n-1}$.
Define $f(x)=x^d$ be a function on $\F_{3^n}$. For $c=-1$, $f(x)$ is PcN.
\end{theorem}

\noindent{\bf Proof.} Since $n,k,d$ are odd and $\frac{3^k+1}{2}\cdot d\equiv\frac{3^n+1}{2}\pmod{3^n-1}$, we get $\gcd(d, 3^n-1)=1$, $\frac{3^k+1}{2}$ is even and
$$(3^k+1)d=l(3^n-1)+2$$ for some odd $l$.

Given $b\in\F_{3^n}$, we will consider the solutions of the equation
\begin{equation}\label{f1}
(x+1)^d+x^d=b
\end{equation}
over $\F_{3^n}$. If $b=0$, we get only one solution $x=1$ of \eqref{f1}.
Assume $x\neq0,-1$ and $b\neq0$. Let $u_{x+1}=(x+1)^d$ and $u_x=-x^d$. Clearly, $u_{x+1}=u_x+b$, $u_{x+1}^{\frac{3^k+1}{2}}=(x+1)\chi(x+1)$
and $u_x^{\frac{3^k+1}{2}}=x\chi(x)$. Set $\frac{u_x}{b}=\eta+\eta^{-1}+1=\frac{(\eta-1)^2}{\eta}$, where $\eta\in\F_{3^{2n}}^*$.
Then $\frac{u_{x+1}}{b}=\frac{u_x}{b}+1=\frac{(\eta+1)^2}{\eta}$ and $\frac{u_{x+1}}{u_x}=(\frac{\eta+1}{\eta-1})^2\in\F_{3^n}$.
We remark that $\eta\not\in\F_3$ for $x\neq0,-1$.
It can be verified that
\begin{equation}\label{f2}
-\frac{(\eta-1)^2 b}{\eta} \chi(x)=x^d \chi(x)=u_x^{\frac{3^k+1}{2}d}=\frac{(\eta-1)^{(3^k+1)d} b^{\frac{3^k+1}{2}d}}{\eta^{\frac{3^k+1}{2}d}}
\end{equation}
and
\begin{equation}\label{f3}
\frac{(\eta+1)^2 b}{\eta} \chi(x+1)=(x+1)^d \chi(x+1)=u_{x+1}^{\frac{3^k+1}{2}d}=\frac{(\eta+1)^{(3^k+1)d} b^{\frac{3^k+1}{2}d}}{\eta^{\frac{3^k+1}{2}d}}.
\end{equation}
From \eqref{f2} and \eqref{f3} we get
\begin{equation}\label{f4}
(\frac{\eta+1}{\eta-1})^{(3^k+1)d-2}=(\frac{\eta+1}{\eta-1})^{l(3^n-1)}=-\frac{\chi(x+1)}{\chi(x)}.
\end{equation}
The solutions of \eqref{f1} can be divided into two cases according to the values of the pair $(\chi(x+1),\chi(x))$.

Case I: $\chi(x+1)=\chi(x)$. In this case, \eqref{f4} can be reduced to $(\frac{\eta+1}{\eta-1})^{l(3^n-1)}=-1$, which implies that $(\frac{\eta+1}{\eta-1})^{l}\not\in\F_{3^n}$.
Recall that $(\frac{\eta+1}{\eta-1})^2\in\F_{3^n}$ and $l$ is odd. We can deduce that $\frac{\eta+1}{\eta-1}\not\in\F_{3^n}$ and $\eta\not\in\F_{3^n}$.
From \eqref{f1} we have
$$(u_x+b)^{\frac{3^k+1}{2}}-u_x^{\frac{3^k+1}{2}}=\chi(x),$$
which implies that $$\frac{(\eta+1)^{3^k+1}}{\eta^{\frac{3^k+1}{2}}}-\frac{(\eta-1)^{3^k+1}}{\eta^{\frac{3^k+1}{2}}}=\chi(x) b^{-\frac{3^k+1}{2}}.$$
It leads to
\begin{equation}\label{f5}
\eta^{\frac{3^k-1}{2}}+\eta^{-\frac{3^k-1}{2}}=-\chi(x) b^{-\frac{3^k+1}{2}}.
\end{equation}
Let $\beta=\eta^{\frac{3^k-1}{2}}$. \eqref{f5} turns to
$$\beta^2+\chi(x) b^{-\frac{3^k+1}{2}}\beta+1=0.$$
Since $k$ is odd, $\gcd(\frac{3^k-1}{2},3^{2n}-1)=1$ and then $\beta\not\in\F_{3^n}$. The above equation has solutions in $\F_{3^{2n}}\setminus\F_{3^n}$ only if
$$\chi(b^{-(3^k+1)}-4)=\chi(1-b^{3^k+1})=-1.$$

If $\chi(x)=1$, we obtain two solutions $\eta_1$ and $\eta_1^{-1}$ of \eqref{f5}. Both solutions lead to
$u_x=(\eta_1+\eta_1^{-1}+1)b$ and $u_{x+1}=(\eta_1+\eta_1^{-1}-1)b$. Since $n$ is odd, we have that $\frac{3^n-1}{2}$ is odd, which implies that
-1 is not a square in $\F_{3^n}$. Therefore, \eqref{f1} has a solution
$x=u_x^{\frac{3^k+1}{2}}=(\eta_1+\eta_1^{-1}+1)^{\frac{3^k+1}{2}} b^{\frac{3^k+1}{2}}$ if $\chi((\eta_1+\eta_1^{-1}+1)b)=-1$ and $\chi((\eta_1+\eta_1^{-1}-1)b)=1$.

If $\chi(x)=-1$, we can derive two solutions $-\eta_1$ and $-\eta_1^{-1}$ from \eqref{f5}.
It follows that $u_x=(-\eta_1-\eta_1^{-1}+1)b$ and $u_{x+1}=-(\eta_1+\eta_1^{-1}+1)b$.
\eqref{f1} has one solution $x=-u_x^{\frac{3^k+1}{2}}=-(-\eta_1-\eta_1^{-1}+1)^{\frac{3^k+1}{2}} b^{\frac{3^k+1}{2}}$
only if $\chi((\eta_1+\eta_1^{-1}+1)b)=1$ and $\chi((\eta_1+\eta_1^{-1}-1)b)=-1$.

Thus, there exists at most one solution of \eqref{f1} in Case I.

Case II: $\chi(x+1)=-\chi(x)$. By \eqref{f4} we have $(\frac{\eta+1}{\eta-1})^{l(3^n-1)}=1$, which implies that $(\frac{\eta+1}{\eta-1})^l\in\F_{3^n}$.
Combining with the known results $(\frac{\eta+1}{\eta-1})^2\in\F_{3^n}$ and $l$ is odd, one has $\frac{\eta+1}{\eta-1}\in\F_{3^n}$ and then $\eta\in\F_{3^n}$. It follows from \eqref{f1} that
$$(u_x+b)^{\frac{3^k+1}{2}}+u_x^{\frac{3^k+1}{2}}=\chi(x+1),$$
which means that $$\frac{(\eta+1)^{3^k+1}}{\eta^{\frac{3^k+1}{2}}}+\frac{(\eta-1)^{3^k+1}}{\eta^{\frac{3^k+1}{2}}}=b^{-\frac{3^k+1}{2}} \chi(x+1).$$
It leads to
\begin{equation}\label{f6}
\eta^{\frac{3^k+1}{2}}+\eta^{-\frac{3^k+1}{2}}=-b^{-\frac{3^k+1}{2}} \chi(x+1).
\end{equation}
Let $\theta=\eta^{(3^k+1)/2}\in\F_{3^n}$. \eqref{f6} turns to
$$\theta^2+b^{-\frac{3^k+1}{2}} \chi(x+1)\theta+1=0.$$
If $1-b^{3^k+1}=0$, we get $b^{(3^k+1)/2}=\pm1$ and $\theta^2\pm\theta+1=0$ from the above equation. It leads to $\theta=\eta^{(3^k+1)/2}=\pm1$. As $\gcd(\frac{3^k+1}{2},3^n-1)=2$,
$\eta^{(3^k+1)/2}=-1$ does not hold in $\F_{3^n}$, and $\eta^{(3^k+1)/2}=1$ has two solutions $\eta=\pm1$. It contradicts the first assumption $\eta\not\in\F_3$.
Therefore, \eqref{f6} has the solutions
$$\theta=\frac{\chi(x+1)\pm \sqrt{1-b^{3^k+1}}}{b^{(3^k+1)/2}}$$
over $\F_{3^n}$ only if $\chi(1-b^{3^k+1})=1$. Moreover, \eqref{f6} can be written as
$$\chi(x+1)\frac{b^{(3^k+1)/2}}{\eta^{(3^k+1)/2}}=-\frac{1}{1+\eta^{3^k+1}},$$
which leads to $$x+1=\chi(x+1)\frac{(\eta+1)^{3^k+1}b^{(3^k+1)/2}}{\eta^{(3^k+1)/2}}=-\frac{(\eta+1)^{3^k+1}}{1+\eta^{3^k+1}}$$
and $$x=-\chi(x+1)\frac{(\eta-1)^{3^k+1}b^{(3^k+1)/2}}{\eta^{(3^k+1)/2}}=\frac{(\eta-1)^{3^k+1}}{1+\eta^{3^k+1}}.$$
Recall that $\eta\in \F_{3^n}$ and $(3^k+1)d=l(3^n-1)+2$. From \eqref{f1} we obtain
$$\begin{array}{lll}
b&=&(-\frac{(\eta+1)^{3^k+1}}{1+\eta^{3^k+1}})^d+(\frac{(\eta-1)^{3^k+1}}{1+\eta^{3^k+1}})^d\\
&=&\frac{(\eta-1)^{(3^k+1)d}-(\eta+1)^{(3^k+1)d}}{(1+\eta^{3^k+1})^d}\\
&=&-\frac{\eta}{(1+\eta^{3^k+1})^d},
\end{array}$$
which implies that
\begin{equation}\label{f7}
\eta=-(1+\eta^{3^k+1})^{d}\cdot b.
\end{equation}
It can be verified that
$$1+\eta^{3^k+1}=1+\theta^2=1+\frac{2-b^{3^k+1}\pm2\chi(x+1) \sqrt{1-b^{3^k+1}}}{b^{3^k+1}}=\frac{-1\pm\sqrt{1-b^{3^k+1}}}{b^{3^k+1}}.$$
Then by \eqref{f7} we get two solutions $$\eta_2=-(\frac{-1+\sqrt{1-b^{3^k+1}}}{b^{3^k+1}})^d \cdot b=\frac{(1-\sqrt{1-b^{3^k+1}})^d}{b}$$ and
$$\eta_3=-(\frac{-1-\sqrt{1-b^{3^k+1}}}{b^{3^k+1}})^d \cdot b=\frac{(1+\sqrt{1-b^{3^k+1}})^d}{b}=\frac{b}{(1-\sqrt{1-b^{3^k+1}})^d}=\eta_2^{-1}.$$
It is obvious that $\eta_2$ and $\eta_3$ give the same value of $x$. So there is at most one solution of \eqref{f1} in this case.

Based on the above discussion, we can conclude that \eqref{f1} has at most one solution in Case I if $\chi(1-b^{3^k+1})=-1$, and has
at most one solution in Case II if $\chi(1-b^{3^k+1})=1$. Note that $x=0$ and -1 are the solutions of \eqref{f1} when $b=1$ and -1 respectively. But when $b=\pm1$,
we have $1-b^{3^k+1}=0$, which means that \eqref{f1} has no solution in Cases I and II. Then by Lemma \ref{lema}, we have that $_{-1}\Delta_f=1$, which completes the proof.
\proved

In Theorem \ref{thmf}, if we replace the condition $d$ is odd with $d$ is even, then we get $_{-1}\Delta_f\leq6$. We omit the proof since it is similar to
the proof of Theorem \ref{thmf}.

\begin{theorem} \label{thmg} Let $d$ be odd. Let $n, k$ be positive integers such that $\gcd(2n,k)=1$ and $\frac{5^k+1}{2}\cdot d\equiv\frac{5^n+1}{2}\pmod{5^n-1}$.
Define $f(x)=x^d$ be a function on $\F_{5^n}$. For $c=-1$, $f(x)$ is PcN.
\end{theorem}

\noindent{\bf Proof.} Since $d$ is odd and $\frac{5^k+1}{2}\cdot d\equiv\frac{5^n+1}{2}\pmod{5^n-1}$, it can be easily checked that $\gcd(\frac{5^k+1}{2}, 5^n-1)=\gcd(d, 5^n-1)=1$ and
$$(5^k+1)d=l(5^n-1)+2$$ for some odd $l$.

For any $b\in\F_{5^n}$, we need to show that the equation
\begin{equation}\label{g1}
(x+1)^d+x^d=b
\end{equation}
has at most one solution over $\F_{5^n}$.  If $b=0$, there is a unique solution $x=2$ of \eqref{g1}.
Assume $x\neq0,-1$ and $b\neq0$. Let $u_{x+1}=(x+1)^d$ and $u_x=-x^d$. Clearly, $u_{x+1}=u_x+b$, $u_{x+1}^{\frac{5^k+1}{2}}=(x+1)\chi(x+1)$
and $u_x^{\frac{5^k+1}{2}}=-x\chi(x)$. Set $\frac{u_x}{b}=\eta+\eta^{-1}+2=\frac{(\eta+1)^2}{\eta}$, where $\eta\in\F_{5^{2n}}^*$.
It leads to $\frac{u_{x+1}}{b}=\frac{u_x}{b}+1=\frac{(\eta-1)^2}{\eta}$ and $\frac{u_{x+1}}{u_x}=(\frac{\eta-1}{\eta+1})^2\in\F_{5^n}$.
As $x\neq0,-1$, then $\eta\not\in\{0,1,-1\}$. It follows that
\begin{equation}\label{g2}
\frac{(\eta+1)^2 b}{\eta} \chi(x)=-x^d \chi(x)=u_x^{\frac{5^k+1}{2}d}=\frac{(\eta+1)^{(5^k+1)d} b^{\frac{5^k+1}{2}d}}{\eta^{\frac{5^k+1}{2}d}}
\end{equation}
and
\begin{equation}\label{g3}
\frac{(\eta-1)^2 b}{\eta} \chi(x+1)=(x+1)^d \chi(x+1)=u_{x+1}^{\frac{5^k+1}{2}d}=\frac{(\eta-1)^{(5^k+1)d} b^{\frac{5^k+1}{2}d}}{\eta^{\frac{5^k+1}{2}d}}.
\end{equation}
From \eqref{g2} and \eqref{g3} we obtain
\begin{equation}\label{g4}
(\frac{\eta+1}{\eta-1})^{(5^k+1)d-2}=(\frac{\eta+1}{\eta-1})^{l(5^n-1)}=\frac{\chi(x)}{\chi(x+1)}.
\end{equation}
We discuss the solutions of \eqref{g1} in two disjoint cases.

Case I: $\chi(x+1)=\chi(x)$. In this case, \eqref{g4} can be reduced to $(\frac{\eta+1}{\eta-1})^{l(5^n-1)}=1$, which implies that $(\frac{\eta+1}{\eta-1})^{l}\in\F_{5^n}$.
As we known, $(\frac{\eta+1}{\eta-1})^2\in\F_{5^n}$ and $l$ is odd. This leads to $\frac{\eta+1}{\eta-1}\in\F_{5^n}$ and $\eta\in\F_{5^n}$.
From \eqref{g1} we have
$$(u_x+b)^{\frac{5^k+1}{2}}+u_x^{\frac{5^k+1}{2}}=\chi(x),$$
which means that $$\frac{(\eta-1)^{5^k+1}}{\eta^{\frac{5^k+1}{2}}}+\frac{(\eta+1)^{5^k+1}}{\eta^{\frac{5^k+1}{2}}}=b^{-\frac{5^k+1}{2}} \chi(x).$$
It leads to
\begin{equation}\label{g5}
\eta^{\frac{5^k+1}{2}}+\eta^{-\frac{5^k+1}{2}}=\frac{1}{2}b^{-\frac{5^k+1}{2}} \chi(x).
\end{equation}
Let $\theta=\eta^{(5^k+1)/2}\in\F_{5^n}$. \eqref{g5} turns to
$$\theta^2+2b^{-\frac{5^k+1}{2}} \chi(x)\theta+1=0.$$
If $1-b^{5^k+1}=0$, we get $b^{(5^k+1)/2}=\pm1$ and $\theta=\eta^{(5^k+1)/2}=\pm1$ from the above equation. Recall that $\gcd(\frac{5^k+1}{2},5^n-1)=1$.
It leads to $\eta=\pm1$, which contradicts the first assumption $\eta\not\in\{0,1,-1\}$.
Hence, \eqref{g5} has the solutions
$$\theta=\frac{-\chi(x)\pm \sqrt{1-b^{5^k+1}}}{b^{(5^k+1)/2}}$$
on $\F_{5^n}$ only if $\chi(1-b^{5^k+1})=1$. Furthermore, \eqref{g5} can be expressed as
$$\chi(x)\frac{b^{(5^k+1)/2}}{\eta^{(5^k+1)/2}}=\frac{1}{2(1+\eta^{5^k+1})},$$
which leads to $$x+1=\chi(x+1)\frac{(\eta-1)^{5^k+1}b^{(5^k+1)/2}}{\eta^{(5^k+1)/2}}=\frac{(\eta-1)^{5^k+1}}{2(1+\eta^{5^k+1})}$$
and $$x=-\chi(x)\frac{(\eta+1)^{5^k+1}b^{(5^k+1)/2}}{\eta^{(5^k+1)/2}}=-\frac{(\eta+1)^{5^k+1}}{2(1+\eta^{5^k+1})}.$$
Since $\eta\in\F_{5^n}$ and $(5^k+1)d=l(5^n-1)+2$, from \eqref{g1} we get
$$\begin{array}{lll}
b&=&(\frac{(\eta-1)^{5^k+1}}{2(1+\eta^{5^k+1})})^d+(-\frac{(\eta+1)^{5^k+1}}{2(1+\eta^{5^k+1})})^d\\
&=&\frac{(\eta-1)^{(5^k+1)d}-(\eta+1)^{(5^k+1)d}}{2^d\cdot(1+\eta^{5^k+1})^d}\\
&=&\frac{\eta}{2^d\cdot(1+\eta^{5^k+1})^d},
\end{array}$$
which implies that
\begin{equation}\label{g6}
\eta=2^{d}\cdot(1+\eta^{5^k+1})^{d}\cdot b.
\end{equation}
Note that
$$1+\eta^{5^k+1}=1+\theta^2=1+\frac{2-b^{5^k+1}\pm2\chi(x) \sqrt{1-b^{5^k+1}}}{b^{5^k+1}}=\frac{2\pm2\sqrt{1-b^{5^k+1}}}{b^{5^k+1}}.$$
Substituting it into \eqref{g6}, we get two solutions $$\eta_1=2^{d}\cdot(\frac{2+2\sqrt{1-b^{5^k+1}}}{b^{5^k+1}})^{d} \cdot b=-\frac{(1+\sqrt{1-b^{5^k+1}})^d}{b}$$ and
$$\eta_2=2^{d}\cdot(\frac{2-2\sqrt{1-b^{5^k+1}}}{b^{5^k+1}})^{d} \cdot b=-\frac{(1-\sqrt{1-b^{5^k+1}})^d}{b}=-\frac{b}{(1+\sqrt{1-b^{5^k+1}})^d}=\eta_1^{-1}.$$
Obviously, $\eta_1$ and $\eta_2$ give the same value of $x$. Hence, \eqref{g1} has at most one solution in this case.

Case II: $\chi(x+1)=-\chi(x)$. By \eqref{g4} we have $(\frac{\eta+1}{\eta-1})^{l(5^n-1)}=-1$. Similar to the proof of Theorem \ref{thmf},
we can get $\eta\not\in\F_{5^n}$, which means that $\eta^{5^n+1}=1$. It follows from \eqref{g1} that
$$(u_x+b)^{\frac{5^k+1}{2}}-u_x^{\frac{5^k+1}{2}}=\chi(x+1),$$
which implies that $$\frac{(\eta-1)^{5^k+1}}{\eta^{\frac{5^k+1}{2}}}-\frac{(\eta+1)^{5^k+1}}{\eta^{\frac{5^k+1}{2}}}=\chi(x+1) b^{-\frac{5^k+1}{2}}.$$
It leads to
\begin{equation}\label{g7}
\eta^{\frac{5^k-1}{2}}+\eta^{-\frac{5^k-1}{2}}=2\chi(x+1) b^{-\frac{5^k+1}{2}}.
\end{equation}
Let $\beta=\eta^{\frac{5^k-1}{2}}$. If $\beta\in\F_{5^n}$, then $\eta^{\frac{5^k-1}{2}}=(\eta^{\frac{5^k-1}{2}})^{5^n}=\eta^{\frac{1-5^k}{2}}$,
which leads to $\eta^{5^k-1}=1$. Since $\gcd(2n,k)=1$, we have $\eta^4=1$. It implies that $\eta\in\F_{5^n}$, which is a contradiction.
Therefore, $\beta\not\in\F_{5^n}$. \eqref{g7} can be written as
$$\beta^2-2\chi(x+1) b^{-\frac{5^k+1}{2}}\beta+1=0.$$
The above equation has the solutions
$$\beta=\frac{\chi(x+1)\pm \sqrt{1-b^{5^k+1}}}{b^{(5^k+1)/2}}$$
in $\F_{5^{2n}}\setminus\F_{5^n}$ only if $\chi(1-b^{5^k+1})=-1$.

Suppose $\eta=\eta_3$ is a solution of \eqref{g7} with $\eta_3^{5^n+1}=1$. It can be verified that
$\gcd(\frac{5^k-1}{2},5^{2n}-1)=2$. As $\chi(x+1)=1$ or -1, the solutions of \eqref{g7} are
$\pm\eta_3$, $\pm\eta_3^{-1}$, $\pm i\eta_3$ and $\pm i\eta_3^{-1}$, where $i^2=-1$.
Note that $(\pm i\eta_3)^{5^n+1}=(\pm i\eta_3^{-1})^{5^n+1}=-1$. It contradicts the first result
$\eta^{5^n+1}=1$. Thus, $\pm i\eta_3$ and $\pm i\eta_3^{-1}$ are not the solutions of \eqref{g7} in this case.

From \eqref{g7} we have that
$$\chi(x+1)\frac{b^{(5^k+1)/2}}{\eta^{(5^k-1)/2}}=\frac{2}{1+\eta^{5^k-1}},$$
which leads to $$x+1=\chi(x+1)\frac{(\eta-1)^{5^k+1}b^{(5^k+1)/2}}{\eta^{(5^k+1)/2}}=\frac{2(\eta-1)^{5^k+1}}{\eta+\eta^{5^k}}$$
and $$x=-\chi(x)\frac{(\eta+1)^{5^k+1}b^{(5^k+1)/2}}{\eta^{(5^k+1)/2}}=\frac{2(\eta+1)^{5^k+1}}{\eta+\eta^{5^k}}.$$
If $\eta=\eta_3$ or $\eta_3^{-1}$, we get
$$(x,x+1)=(\frac{2(\eta_3+1)^{5^k+1}}{\eta_3+\eta_3^{5^k}},\frac{2(\eta_3-1)^{5^k+1}}{\eta_3+\eta_3^{5^k}}).$$
Substituting it into \eqref{g1} gives
\begin{equation}\label{g8}
\big(\frac{2(\eta_3-1)^{5^k+1}}{\eta_3+\eta_3^{5^k}} \big)^d+\big(\frac{2(\eta_3+1)^{5^k+1}}{\eta_3+\eta_3^{5^k}})^d=b.
\end{equation}
If $\eta=-\eta_3$ or $-\eta_3^{-1}$, we obtain
$$(x,x+1)=(-\frac{2(\eta_3-1)^{5^k+1}}{\eta_3+\eta_3^{5^k}},-\frac{2(\eta_3+1)^{5^k+1}}{\eta_3+\eta_3^{5^k}}).$$
Then by \eqref{g1} we have that
\begin{equation}\label{g9}
-\big(\frac{2(\eta_3+1)^{5^k+1}}{\eta_3+\eta_3^{5^k}} \big)^d-\big(\frac{2(\eta_3-1)^{5^k+1}}{\eta_3+\eta_3^{5^k}})^d=b
\end{equation}
since $d$ is odd. Note that \eqref{g8} and \eqref{g9} hold simultaneously only if $b=0$. Since $b\neq0$, only one of them fits \eqref{g1}.
Therefore, there exists at most one solution of \eqref{g1} if $\eta=\eta_3, \eta_3^{-1}$ or $\eta=-\eta_3, -\eta_3^{-1}$.

Based on the above discussion, we can conclude that \eqref{g1} has at most one solution in Case I if $\chi(1-b^{5^k+1})=1$, and has
at most two solutions in Case II if $\chi(1-b^{5^k+1})=-1$. Note that $x=0$ and -1 are solutions of \eqref{g1} when $b=1$ and -1 respectively.
If $b=\pm1$, then $1-b^{5^k+1}=0$, which implies that \eqref{g1} has no solution in Cases I and II.
Then we get the desired result by Lemma \ref{lema}.
\proved

\begin{theorem} \label{thma} Let $n, k$ be positive integers such that $p^n\equiv3 \pmod{4}$ and $d(p^k+1)\equiv\frac{p^n+1}{2} \pmod{p^n-1}$.
Let $f(x)=x^d$ be a function over $\F_{p^n}$. If $c=0$, then $_{c}\Delta_f=1$ when
$d$ is odd, and $_{c}\Delta_f=2$ when $d$ is even. If $c=1$, then $_{c}\Delta_f\leq6$ when
$d$ is odd, and $_{c}\Delta_f\leq3$ when $d$ is even. And if $c=-1$, then $_{c}\Delta_f\leq3$ when
$d$ is odd, and $_{c}\Delta_f\leq6$ when $d$ is even.
\end{theorem}

\noindent{\bf Proof.} Since $d(p^k+1)\equiv\frac{p^n+1}{2} \pmod{p^n-1}$ and $p^n\equiv3 \pmod{4}$, it can be verified that
$$\gcd(d(p^k+1),p^n-1)=\gcd(\frac{p^n+1}{2},p^n-1)=2$$
and $\gcd(p^k+1,p^n-1)=2$. If $d$ is odd, then $\gcd(d,p^n-1)=1$. If $d$ is even, then $\gcd(d,p^n-1)=2$.
By Lemma \ref{lema}, we have that $_{0}\Delta_f=1$ when
$d$ is odd, and $_{0}\Delta_f=2$ when $d$ is even.

Given $b\in\F_{p^n}$, we consider the solutions of
\begin{equation}\label{a1}
(x+1)^d-cx^d=b
\end{equation}
over $\F_{p^n}$ for $c=\pm1$. We remark that $\frac{p^n-1}{2}$ is odd, which implies that -1 is not a square over $\F_{p^n}$.
When $b=0$, for odd $d$, \eqref{a1} has no solution if $c=1$, and one solution $-\frac{1}{2}$ if $c=-1$.
Similarly, for even $d$, \eqref{a1} has a unique solution $-\frac{1}{2}$ if $c=1$, and no solution if $c=-1$.

Assume $x\neq0,-1$ and $b\neq0$. Let $u=(x+1)^d$ and $v=cx^d$.
It leads to $u=v+b$, $u^{p^k+1}=(v+b)^{p^k+1}=(x+1)\chi(x+1)$ and $v^{p^k+1}=x\chi(x)$.
In the following, if $v=v_i$ is defined, then $u=u_i$ is automatically defined by the above conditions.   
We can divide the solutions of \eqref{a1} into the following four cases according to
the values of $\chi(x+1)$ and $\chi(x)$.

Case I: $(\chi(x+1),\chi(x))=(1,1)$. Based on the value of the pair $(\chi(x+1),\chi(x))$, we get $(v+b)^{p^k+1}-v^{p^k+1}=1$, which can be reduced to
\begin{equation}\label{a2}
bv^{p^k}+b^{p^k}v=1-b^{p^k+1}.
\end{equation}
Recall that -1 is not a square over $\F_{p^n}$. Then $-b^{p^k-1}$ is not a $(p^k-1)$th power in $\F_{p^n}$.
Applying Lemma \ref{lemc}, \eqref{a2} has exactly one solution over $\F_{p^n}$, which can be denoted as $v_1$.
It leads to $(x, x+1)=(v_1^{p^k+1},v_1^{p^k+1}+1)$.

Case II: $(\chi(x+1),\chi(x))=(-1,-1)$. In this case, we have $(v+b)^{p^k+1}-v^{p^k+1}=-1$, which leads to
\begin{equation}\label{a3}
bv^{p^k}+b^{p^k}v=-1-b^{p^k+1}.
\end{equation}
Similarly, there is only one solution $v=v_2$ over $\F_{p^n}$. Correspondingly, we get $u=u_2$ satisfying that $u_2=v_2+b=-v_1$.
It can be checked that $x+1=-u_2^{p^k+1}=-v_1^{p^k+1}$ and $x=-v_1^{p^k+1}-1$.

Case III: $(\chi(x+1),\chi(x))=(1,-1)$. Given the value of the pair $(\chi(x+1),\chi(x))$, we obtain $(v+b)^{p^k+1}+v^{p^k+1}=1$, which implies that
\begin{equation}\label{a4}
(v+\frac{b}{2})^{p^k+1}=\frac{2-b^{p^k+1}}{4}.
\end{equation}
There are two solution $v_3$ and $v_4$ over $\F_{p^n}$ with $v_3=-v_4-b=-u_4$. It follows that
$(x,x+1)=(-v_3^{p^k+1},-v_3^{p^k+1}+1)$ or $(v_3^{p^k+1}-1,v_3^{p^k+1})$.

Case IV: $(\chi(x+1),\chi(x))=(-1,1)$. In this case, we have $(v+b)^{p^k+1}+v^{p^k+1}=-1$, which means that
\begin{equation}\label{a5}
(v+\frac{b}{2})^{p^k+1}=\frac{-2-b^{p^k+1}}{4}.
\end{equation}
There are two solution $v_5$ and $v_6$ over $\F_{p^n}$ with $v_5=-v_6-b=-u_6$. Then we can deduce that
$(x,x+1)=(v_5^{p^k+1},v_5^{p^k+1}+1)$ or $(-v_5^{p^k+1}-1,-v_5^{p^k+1})$.

Suppose $d$ is even. Next we consider the case of $c=1$. It is easy to see that 
$$\chi(u)=\chi((x+1)^d)=1\; {\rm and}\; \chi(v)=\chi(x^d)=1.$$
From Case I we get $\chi(v_1)=1$. From Case II we have $\chi(u_2)=\chi(-v_1)=1$. It leads to $\chi(v_1)=-1$, which is a contradiction.
Therefore, there is at most one solution of \eqref{a1} in Cases I and II. In Case III, if $v_3$ and $v_4$ are both the solutions of \eqref{a1}, then we have $\chi(v_3)=\chi(-u_4)$. It contradicts
that -1 is not a square in $\F_{p^n}$. Therefore, there is at most one solution of \eqref{a1} in Case III. Similarly, we can show that
\eqref{a1} has at most one solution in Case IV. Hence, we can conclude that \eqref{a1} has at most three solutions in Cases I, II, III and IV. For the case of
$c=-1$, we obviously have that \eqref{a1} has at most six solutions in the above four cases.

Suppose $d$ is odd. For $c=1$, it can be easily checked that there are at most six solutions of \eqref{a1} in Cases I-IV. Now we discuss the case of $c=-1$.
Clearly, $$\chi(u)=\chi((x+1)^d)=\chi(x+1)\; {\rm and}\; \chi(v)=\chi(-x^d)=-\chi(x).$$ 
In Case I we get $\chi(v_1)=-\chi(x)=-1$. And in Case II we have $\chi(-v_1)=\chi(u_2)=\chi(x+1)=-1$, which implies that $\chi(v_1)=1$. It leads to a contradiction.
Hence, there exists at most one solution of \eqref{a1} in Cases I and II. Assume that $v_3$ and $v_4$ are both the solutions of \eqref{a1} in Case III.
Then we have $\chi(v_3)=\chi(-u_4)$, where $\chi(v_3)=-\chi(x)=1$ and $\chi(-u_4)=-\chi(x+1)=-1$. This leads to a contradiction.
Thus, we obtain at most one solution of \eqref{a1} in Case III. Similarly, we find that
\eqref{a1} has at most one solution in Case IV. We summary that there exist at most three solutions of \eqref{a1} in the mentioned four cases.

Note that $x=0$ and $-1$ are the solutions of \eqref{a1} when $b=1$ and $(-1)^{d+1}c$ respectively.
In the sequel, for $b=\pm1$, we analyze the solutions of \eqref{a1} in $\{0,-1\}$, Cases I, II, III and IV.   
Firstly, we consider the case of $b=1$. If $c=1$ and $d$ is odd, it can be checked that \eqref{a1} has at most four solutions 0, -1, $v_5^{p^k+1}$ and $-v_5^{p^k+1}-1$.
If $c=1$ and $d$ is even, \eqref{a1} has at most two solutions (0 and one solution in Case IV).
If $c=-1$ and $d$ is odd, \eqref{a1} has at most two solutions (0 and one solution in Case IV).
And if $c=-1$ and $d$ is even, \eqref{a1} has at most four solutions 0, -1, $v_5^{p^k+1}$ and $-v_5^{p^k+1}-1$.

Secondly, we consider the case of $b=-1$. If $c=1$ and $d$ is odd, \eqref{a1} has at most two solutions $v_5^{p^k+1}$ and $-v_5^{p^k+1}-1$.
If $c=1$ and $d$ is even, \eqref{a1} has at most two solutions (-1 and one solution in Case IV).
If $c=-1$ and $d$ is odd, \eqref{a1} has at most two solutions (-1 and one solution in Case IV).
And if $c=-1$ and $d$ is even, \eqref{a1} has at most two solutions $v_5^{p^k+1}$ and $-v_5^{p^k+1}-1$.

Employing Lemma \ref{lema}, we can conclude the desired results by the above discussions.
\proved

\begin{theorem} \label{thmb}
Let $p>3$ and $c\in\F_{p^n}$ with $c\neq1$. For a function $f(x)=x^3$ defined on $\F_{p^n}$, there is $_{c}\Delta_f\leq3$.
\end{theorem}

\noindent{\bf Proof.} Set $b\in\F_{p^n}$, we look at the equation $(x+1)^3-cx^3=b$.
If $c=0$, then $_{c}\Delta_f=\gcd(3,p^n-1)$. It follows that $_{0}\Delta_f=1$ when $\gcd(3,p^n-1)=1$,
and $_{0}\Delta_f=3$ when $\gcd(3,p^n-1)=3$.
If $c\neq0,1$, then we have $$(1-c)x^3+3x^2+3x+1-b=0,$$ which implies that
$$x^3+\frac{3}{1-c}x^2+\frac{3}{1-c}x+\frac{1-b}{1-c}=0.$$
Let $z=x+\frac{3}{1-c}$. The above equation leads to
$$z^3-\frac{3c}{(1-c)^2}z+\frac{3c-1}{(1-c)^3}+\frac{1-b}{1-c}=0,$$
which has at most three solutions over $\F_{p^n}$. Therefore, $_{c}\Delta_f\leq3$ for any $c\in\F_{p^n}\setminus\{1\}$.
\proved

The $c$-differential uniformity of $x^{\frac{2p^n-1}{3}}$ is shown in \cite{YMZ}. Note that $x^3$ is the inverse mapping of $x^{\frac{2p^n-1}{3}}$ over $\F_{p^n}$
only if $p^n\equiv2 \pmod{3}$. Therefore, the result of Theorem \ref{thmb} is new.

\begin{theorem} \label{thmd} Let $f(x)=x^d$ be a function defined on $\F_{p^n}$, where $d=\frac{p^n+1}{4}+\frac{p^n-1}{2}$ if $p^n\equiv7 \pmod{8}$, and $d=\frac{p^n+1}{4}$ if $p^n\equiv3 \pmod{8}$.
For $c=-1$, $_{c}\Delta_f\leq3$.
\end{theorem}

\noindent{\bf Proof.} Under the conditions of the theorem, it can be checked that $2d\equiv\frac{p^n+1}{2} \pmod{p^n-1}$,
$\gcd(d,p^n-1)=1$, both $d$ and $\frac{p^n-1}{2}$ are odd. Then -1 is not a square in $\F_{p^n}$. For $c=-1$, we explore the solutions of equation
\begin{equation}\label{d1}
(x+1)^{d}+x^{d}=b
\end{equation}
over $\F_{p^n}$ for $b\in\F_{p^n}$. Clearly, \eqref{d1} has a unique solution $x=-\frac{1}{2}$ when $b=0$.

Assume $b\neq0$ and $x\neq0, -1$. Let $u_{x+1}=(x+1)^d$ and $u_x=x^d$. Then $u_{x+1}=b-u_x$, which leads to
$$u_{x+1}^2=(b-u_x)^2=u_x^2-2bu_x+b^2.$$
Note that $u_{x+1}^2=(x+1)\chi(x+1)$ and $u_x^2=x\chi(x)$. It follows that
\begin{equation}\label{d2}
(x+1)\chi(x+1)=x\chi(x)-2bu_x+b^2.
\end{equation}
In the following, we analyze the solutions of \eqref{d2} in four disjoint cases.

Case I: $(\chi(x+1),\chi(x))=(1,1)$. In this case, we get $x+1=x-2bu_x+b^2$, which implies that $u_x=\frac{b^2-1}{2b}$, $u_{x+1}=b-u_x=\frac{b^2+1}{2b}$ and $x=(\frac{b^2-1}{2b})^2$.
Recall that $d$ is odd. $x$ a solution in Case I only if $\chi(\frac{b^2-1}{2b})=\chi(\frac{b^2+1}{2b})=1$.

Case II: $(\chi(x+1),\chi(x))=(-1,-1)$. \eqref{d2} becomes $-x-1=-x-2bu_x+b^2$. Then we have $u_x=\frac{b^2+1}{2b}$, $u_{x+1}=b-u_x=\frac{b^2-1}{2b}$ and $x=-(\frac{b^2+1}{2b})^2$,
which holds only if $\chi(\frac{b^2-1}{2b})=\chi(\frac{b^2+1}{2b})=-1$. 

It is easy to see that there exists at most one solution of \eqref{d2} in Cases I and II.

Case III: $(\chi(x+1),\chi(x))=(1,-1)$. Similarly, we can get $x+1=-x-2bu_x+b^2$ from \eqref{d2}. This implies that
\begin{equation}\label{d3}
u_x^2-bu_x+\frac{b^2-1}{2}=0
\end{equation}
since $u_x^2=x\chi(x)=-x$. \eqref{d3} has at most two solutions $u_x=\frac{b+\sqrt{2-b^2}}{2}$ and $u_x^*=\frac{b-\sqrt{2-b^2}}{2}$,
which leads to $u_{x+1}=\frac{b-\sqrt{2-b^2}}{2}$ and $u_{x+1}^*=\frac{b+\sqrt{2-b^2}}{2}$ correspondingly.
If $u_x=\frac{b+\sqrt{2-b^2}}{2}$ is a solution of \eqref{d2}, then $\chi(\frac{b+\sqrt{2-b^2}}{2})=-1$.
If $u_x^*=\frac{b-\sqrt{2-b^2}}{2}$ is a solution of \eqref{d2}, then $\chi(u_{x+1}^*)=\chi(\frac{b+\sqrt{2-b^2}}{2})=1$.
Hence, $u_x$ and $u_x^*$ are not the solutions of \eqref{d2} simultaneously. Then we get at most one solution $x$ of \eqref{d2}.

Case IV: $(\chi(x+1),\chi(x))=(-1,1)$. In this case, \eqref{d2} becomes $-x-1=x-2bu_x+b^2$.
As $u_x^2=x\chi(x)=x$, it follows that
\begin{equation}\label{d4}
u_x^2-bu_x+\frac{b^2+1}{2}=0.
\end{equation}
It can be checked that \eqref{d4} has at most two solutions $u_x=\frac{b+\sqrt{-2-b^2}}{2}$ and $u_x^*=\frac{b-\sqrt{-2-b^2}}{2}$,
which leads to $u_{x+1}=\frac{b-\sqrt{-2-b^2}}{2}$ and $u_{x+1}^*=\frac{b+\sqrt{-2-b^2}}{2}$ correspondingly.
Similar to the discussion of Case III, we can show that \eqref{d2} has at most one solution in this case.

We remark that 0 and -1 are solutions of \eqref{d1} while $b=1$ and -1 respectively. If $b=1$, it can be verified that
\eqref{d1} has at most two solutions (0 and one solution in Case IV). If $b=-1$, we can also get that \eqref{d1} has at most two solutions (0 and one solution in Case IV).
And if $b\neq0,\pm1$, \eqref{d1} has at most three solutions in Cases I, II, III and IV. Then we get the conclusion $_{-1}\Delta_f\leq3$ by Lemma \ref{lema}.
\proved

It can be seen in Theorem \ref{thm1} that $g(x)=x^d$ is APN over $\F_{p^n}$, where $d=\frac{p^n+1}{4}$ if $p^n\equiv7 \pmod{8}$, and $d=\frac{p^n+1}{4}+\frac{p^n-1}{2}$ if $p^n\equiv3 \pmod{8}$.
Similar to the proof of Theorem \ref{thmd}, we can prove that $_{-1}\Delta_g\leq6$ since $d$ is even. The proof is omitted here.

\begin{theorem} \label{thme} Let $n, k$ be positive integers such that $\frac{n}{\gcd(n,k)}$ is odd. Let $f(x)=x^{\frac{p^n-1}{2}+p^k+1}$ be a function defined on $\F_{p^n}$. If $p^n\equiv3 \pmod{4}$,
then $_{-1}\Delta_f\leq3$ and $_{1}\Delta_f\leq6$. If $p^n\equiv1 \pmod{4}$, then $_{-1}\Delta_f\leq6$ and $_{1}\Delta_f\leq3$.
\end{theorem}

\noindent{\bf Proof.} Without loss of generality, we just prove the case of $p^n\equiv3 \pmod{4}$. 
Since $\frac{n}{\gcd(n,k)}$ is odd, by Lemma \ref{lemb} we get $\gcd(p^k+1,p^n-1)=2$.
Then $\frac{p^n-1}{2}+p^k+1$ is odd and $\gcd(\frac{p^n-1}{2}+p^k+1,p^n-1)=1$.

For any $b\in\F_{p^n}$, we need to compute the maximal number of the solutions of
\begin{equation}\label{e1}
(x+1)^{\frac{p^n-1}{2}+p^k+1}-cx^{\frac{p^n-1}{2}+p^k+1}=b
\end{equation}
over $\F_{p^n}$ for $c=\pm1$. Assume $x\neq0,-1$. The solutions of \eqref{e1} can be divided into four cases according to the values of the pair $(\chi(x+1),\chi(x))$.

Case I: $(\chi(x+1),\chi(x))=(1,1)$. From \eqref{e1} we get $(x+1)^{p^k+1}-cx^{p^k+1}=b$.
If $c=1$, we have $x^{p^k}+x=b-1$. Since $p^n\equiv3 \pmod{4}$, $\frac{p^n-1}{2}$ is odd, and then -1 is not a square in $\F_{p^n}$.
It follows from Lemma \ref{lemc} that \eqref{e1} has just one solution $x_1$ over $\F_{p^n}$.
If $c=-1$, we obtain $2x^{p^k+1}+x^{p^k}+x=b-1$, which implies that
$$(x+\frac{1}{2})^{p^k+1}=\frac{2b-1}{4}.$$
Recall that $\gcd(p^k+1,p^n-1)=2$. The above equation has two solutions $x_2=-\frac{1}{2}+\theta$ and $x_3=-\frac{1}{2}-\theta$, where $\theta^{p^k+1}=\frac{2b-1}{4}$.
It is easy to see that $x_2+1=-x_3$. 
Therefore, $\chi(x_2+1)=-\chi(x_3)$, which implies that $x_2$ and $x_3$ are not the solutions of \eqref{e1} simultaneously.

Case II: $(\chi(x+1),\chi(x))=(-1,-1)$. In this case, we have $-(x+1)^{p^k+1}+cx^{p^k+1}=b$.
If $c=1$, we obtain $x^{p^k}+x=-b-1$. It leads to one solution $x_4$ of \eqref{e1}.
If $c=-1$, we get $2x^{p^k+1}+x^{p^k}+x=-b-1$, which implies that
$$(x+\frac{1}{2})^{p^k+1}=\frac{-2b-1}{4}.$$
Similarly, we can get two solutions $x_5=-\frac{1}{2}+\beta$ and $x_6=-\frac{1}{2}-\beta$, where $\beta^{p^k+1}=\frac{-2b-1}{4}$.
It can be checked that $x_5+1=-x_6$, and there is at most one solution $x_5$ or $x_6$ of \eqref{e1}.

Case III: $(\chi(x+1),\chi(x))=(1,-1)$. \eqref{e1} leads to $(x+1)^{p^k+1}+cx^{p^k+1}=b$ in this case. If $c=1$, we have
$$(x+\frac{1}{2})^{p^k+1}=\frac{2b-1}{4}.$$
Just like the proof in Case I, the above equation has two solutions $x_2=-\frac{1}{2}+\theta$ and $x_3=-\frac{1}{2}-\theta$, where $\theta^{p^k+1}=\frac{2b-1}{4}$.
If $c=-1$, then $x^{p^k}+x=b-1$, which leads to one solution $x_1$ of \eqref{e1}.

Case IV: $(\chi(x+1),\chi(x))=(-1,1)$. \eqref{e1} becomes $(x+1)^{p^k+1}+cx^{p^k+1}=-b$.
If $c=1$, we have
$$(x+\frac{1}{2})^{p^k+1}=\frac{-2b-1}{4}.$$
As shown in Case II, \eqref{e1} has at most two solutions $x_5=-\frac{1}{2}+\beta$ and $x_6=-\frac{1}{2}-\beta$, where $\beta^{p^k+1}=\frac{-2b-1}{4}$.
If $c=-1$, then $x^{p^k}+x=-b-1$, which leads to one solution $x_4$ of \eqref{e1}.

It can be verified that $x_1+1=-x_4$ and $\chi(x_1+1)=-\chi(x_4)$. Therefore, \eqref{e1} has at most
six solutions ($x_1$, $x_4$, two solutions in Case III and two solutions in Case IV) if $c=1$. And if $c=-1$, \eqref{e1} has at most
three solutions ($x_1$ or $x_4$, one solution in Case I and one solution in Case II).

For $c=1$, $x=0,-1$ are the solutions of \eqref{e1} when $b=1$. If $b=1$, we can easily check that there are at most four solutions (0, -1 and two solutions in Case IV) of \eqref{e1}.
For $c=-1$, $x=0$ and -1 are solutions of \eqref{e1} when $b=1$ and -1 respectively. If $b=1$, \eqref{e1} has at most two solutions (0 and one solution in Case II).
And if $b=-1$, \eqref{e1} has at most two solutions (-1 and one solution in Case I). Then by Lemma \ref{lema}, we can conclude that $_{-1}\Delta_f\leq3$ and $_{1}\Delta_f\leq6$.
\proved

It is well known that the power function $x^{p^k+1}$ is PN over $\F_{p^n}$ when $\frac{n}{\gcd(n,k)}$ is odd \cite{HPRS}. Theorem \ref{thme} shows 
the $c$-differential uniformity of a variant of this PN power function for $c=\pm1$.  

\section{Concluding remarks}

In this paper, we explore the $c$-differential uniformity of some known APN power functions in odd characteristic.
By employing the quadratic character of $\mathbb{F}_{p^n}^*$, we obtain six classes of power functions with low $c$-differential uniformity.
Especially, two of them are PcN for $c=-1$. We believe that one can find more functions with low $c$-differential uniformity ($c\neq1$)
from functions with the usual low differential uniformity.


\begin{thebibliography}{99}

\bibitem{AGW} A. Akbary, D. Ghioca and Q. Wang, {\em On constructing permutations of finite
fields}, Finite Fields Appl. 17(1) (2011) 51-67.

\bibitem{BC} D. Bartoli and M. Calderini, {\em On construction and (non)existence of $c$-almost perfect nonlinear functions}, http://arxiv.org/abs/2008.03953.

\bibitem {BCJW} N. Borisov, M. Chew, R. Johnson and D. Wagner, {\em Multiplicative Differentials}, In: Daemen J., Rijmen V. (eds) Fast Software Encryption.
FSE 2002. Lecture Notes in Computer Science, vol 2365. Springer, Berlin, Heidelberg, 2002.

\bibitem{BS} E. Biham and A. Shamir, {\em Differential cryptanalysis of DES-like cryptosystems}, In Alfred Menezes and Scott A. Vanstone, editors,
Advances in Cryptology-CRYPTO' 90, 10th Annual International Cryptology Conference, Santa Barbara, California, USA, August
11-15, 1990, Proceedings, volume 537 of Lecture Notes in Computer Science, pages 2-21. Springer, 1990.

\bibitem{BT} D. Bartoli and M. Timpanella, {\em On a generalization of planar functions}, J. Algebr. Comb., DOI:https://doi.org/10.1007/s10801-019-
00899-2, 2019.

\bibitem {CCZ} C. Carlet, P. Charpin and V. Zinoviev, {\em Codes, bent functions and permutations suitable for DES-like cryptosystems}, Des. Codes Cryptogr. 15(2) (1998) 125-156.

\bibitem{CM} R.S. Coulter and R.W. Mathews, {\em Planar functions and planes of Lenz-Barlotti class II}, Des. Codes Cryptogr. 10 (1997) 167¨C184.

\bibitem {Dem} U. Dempwolff, {\em CCZ equivalence of power functions}, Des. Codes Cryptogr. 86(3) (2018) 665-692.

\bibitem {DMMP} H. Dobbertin, D. Mills, E.N. M¨¹ller, A. Pott and W. Willems, {\em APN functions in odd characteristic}, Discrete Math. 267(1-3) (2003) 95-112.

\bibitem {EFRST} P. Ellingsen, P. Felke, C. Riera, P. St\u{a}nic\u{a} and A. Tkachenko, {\em $C$-differentials, multiplicative uniformity and (almost) perfect $c$-nonlinearity},
IEEE Trans. Inform. Theory, 2020. To appear.

\bibitem {HPRS} S. U. Hasan, M. Pal, C. Riera and P. St\u{a}nic\u{a}, {\em On the c-differential uniformity of certain maps over finite fields}, https://arxiv.org/abs/2004.09436.

\bibitem{HRS} T. Helleseth, C. Rong and D. Sandberg, {\em New families of almost perfect nonlinear power mappings}, IEEE Trans. Inf. Theory 45(2) (1999) 474-485.

\bibitem{HMSY} X. Hou, G.L. Mullen, J.A. Sellers and J.L. Yucas, {\em Reversed Dickson polynomials over finite fields}, Finite Fields Appl. 15(3) (2009) 748-773.

\bibitem{Led} E. Leducq, {\em New families of APN functions in characteristic 3 or 5}, Contemporary Mathematics 574 (2012) 115¨C123.

\bibitem{MQ} S. Mesnager and L. Qu, {\em On two-to-one mappings over finite fields}, IEEE Trans. Inf. Theory 65(12) (2019) 7884-7895.

\bibitem{RS} C. Riera and P. St\u{a}nic\u{a}, {\em Investigations on $c$-(almost) perfect nonlinear functions}, http://arxiv.org/abs/2004.02245.

\bibitem{SGGRT} P. St\u{a}nic\u{a}, S. Gangopadhyy, A. Geay, C. Riera and A. Tkachenko, {\em C-differential bent functions and perfect nonlinearity}, http://arxiv.org/abs/2006.12535.

\bibitem{YMZ} H. Yan, S. Mesnager and Z. Zhou, {\em Power functions over finite fields with low $c$-differential uniformity}, https://arxiv.org/abs/2003.13019.

\bibitem{ZW} Z. Zha and X. Wang, {\em Power functions with low uniformity on odd characteristic finite fields}, Sci. China Math. 53(8) (2010) 1931¨C1940.

\bibitem{ZW1} Z. Zha and X. Wang, {\em Almost perfect nonlinear power functions in odd characteristic}, IEEE Trans. Inf. Theory 57(7) (2011) 4826-4832.
\end{thebibliography}
\end{document}